\documentclass[conference]{IEEEtran}
\usepackage{url}
\usepackage{hyperref}
\usepackage{xspace}
\usepackage{array}
\usepackage{multirow}
\usepackage[dvipsnames]{xcolor}
\newcommand{\etal}{{\em et al.}\xspace}
\newcommand{\ie}{{\em i.e.}\xspace}
\newcommand{\eg}{{\em e.g.}\xspace}
\newcommand{\BfPara}[1]{{\vspace{1mm}\noindent\bf#1.}\xspace}

\hypersetup{
    pdfpagemode=pagewidth,
    plainpages=false,
    colorlinks,
    urlcolor=OliveGreen,
    linkcolor=OrangeRed,
    citecolor=Cerulean,
    bookmarksnumbered
}

\title{The Landscape of Domain Name Typosquatting: \\ Techniques and Countermeasures}
\date{\today}

\author{
\IEEEauthorblockN{Jeffrey Spaulding}
\IEEEauthorblockA{SUNY Buffalo\\
Buffalo, NY, USA\\
Email: jjspauld@buffalo.edu}
\and
\IEEEauthorblockN{Shambhu Upadhyaya}
\IEEEauthorblockA{SUNY Buffalo\\
Buffalo, NY, USA\\
Email: shambhu@buffalo.edu}
\and
\IEEEauthorblockN{Aziz Mohaisen}
\IEEEauthorblockA{SUNY Buffalo\\
Buffalo, NY, USA\\
Email: mohaisen@buffalo.edu}}

\begin{document}

\maketitle

\begin{abstract}
With more than 294 million registered domain names as of late 2015, the domain name ecosystem has evolved to become a cornerstone for the operation of the Internet. Domain names today serve everyone, from individuals for their online presence to big brands for their business operations. Such ecosystem that facilitated legitimate business and personal uses has also fostered ``creative'' cases of misuse, including phishing, spam, hit and traffic stealing, online scams, among others. As a first step towards this misuse, the registration of a legitimately-looking domain is often required. For that, domain typosquatting provides a great avenue to cybercriminals to conduct their crimes. 

In this paper, we review the landscape of domain name typosquatting, highlighting models and advanced techniques for typosquatted domain names generation, models for their monetization, and the existing literature on countermeasures. We further highlight potential fruitful directions on technical countermeasures that are lacking in the literature. 
 
\end{abstract}

%\ccsdesc[200]{Computer Communication Networks} 

%
% End generated code
%

%
%  Use this command to print the description
%
%\printccsdesc

% We no longer use \terms command
%\terms{Theory}

\BfPara{Keywords} Domain Names, Typosquatting, Defenses.

\section{Introduction}\label{sec:introduction}
Ever since the process of the domain name registration began in the 1990's, cybercriminals have seized the opportunity to profit on the backs of others by misusing such process in so many ways~\cite{Mohaisen15,WestM14,ThomasM14b}. As Internet commerce quickly rose and more companies began registering domain names to get a foothold on the action, certain individuals realized that they could preemptively register these domain names on a first-come first-serve basis. These so called ``cybersquatters'' would purchase domain names in the hopes of selling them back to companies and trademark owners for a substantial profit. As these popular domain names attracted more users to their websites, it was not long before cybercriminals recognized that people often made mistakes when typing URLs into their browsers--thus sparking a new form of domain name exploitation called {\em typosquatting}. 

In this paper, we survey the landscape of {\em typosquatting}, which is the deliberate registration of a domain name that uses typographical variants of other target domain names. Typically, these variant domain names are generated in such a way as to exploit common typographical errors made by users that manually type URLs into web browsers. For example, the popular social networking site Facebook was the target of several typosquatters who registered domain names such as \url{www.fagebook.com} and \url{www.facewbook.com}~\cite{Roth2013}. Unfortunately for these typosquatters, they were ordered to transfer over their domain names and pay Facebook up to \$1.34 million in damages.

As we will discuss in subsequent sections, other forms of domain squatting have emerged that not only uses common typographical mistakes, but employs the use of (among others): visually-similar letters~\cite{Holgers2006}, similar-sounding words~\cite{Nikiforakis2014} or the exploitation of hardware errors that store domain names~\cite{Nikiforakis2013}. In addition, it has been shown that not only are the most popular domain names targeted by typosquatters, the ``long tail'' of the popularity distribution has also come under their sights as potential targets for exploitation~\cite{Szurdi2014}. By providing a comprehensive treatment of typosquatting, we hope that this paper will catapult research on mitigating this problem.

\BfPara{Organization} In \textsection\ref{sec:anatomy} we review the anatomy of typosquatting. In \textsection\ref{sec:monetization} we review the monetization techniques of typosquatting. In \textsection\ref{sec:countermeasures} we review the countermeasures to typosquatting. In \textsection\ref{sec:conclusion} we provide concluding remarks and open directions.

\section{Typosquatting Anatomy}\label{sec:anatomy}
While typosquatting as a phenomenon is perhaps known for many years, the term itself has been in use for almost two decades. Several studies have been conducted to understand models of typosquatting, including advanced techniques and squatted domains features. In the following, we briefly review the historical background of typosquatting, and follow it by a technical anatomy of models, techniques and features.
\vspace{-1mm}
\subsection{Historical Background}
The term {\em typosquatter} may have been coined as far back as 1998 by R. C. Cumbow in The New York Law Journal (NYLJ)~\cite{Gilwit2003}, who was one of the first to write about this new trend of cybersquatting. One of the first large-scale studies on typosquatting was conducted in 2003 by Edelman~\cite{Edelman}, who located more than 8,800 registered domains that were minor typographical variations of popular domain names. Surprisingly, most of these domain names were traced back to one individual, John Zuccarini, who often redirected users to sexually-explicit content and even used nefarious tactics to ``mousetrap'' these users from leaving these sites (\eg blocking the ordinary operation of a browser's Back and Close commands). Some of these typosquatted domain names went so far as to target websites frequentDly visited by children, such as {\tt disenystore.com} (a typo on {\tt disneystore.com}) which redirected to a website with sexually-explicit content.

\subsection{Identifying Typo Domains: Models}\label{sec:models}
Prior experiments conducted on the subject of typosquatting typically began their data collection phase by first identifying a set of domain names and then generating a list of possible typo variations on those domain names. Often these experiments used a subset of the top-ranking domain names according to some domain ranking websites, such as Alexa. The rationale of using such domains is that typosquatters will naturally target the most popular domain names to increase the chances of obtaining unsuspecting visitors. Table~\ref{tab:tab1} summarizes these several approaches of which authoritative domains they studied, the number of possible typosquatted domains they generated, and what percentage of them were active (\ie resolved to an IP address hosting a website). 
In the following section, we describe the models that generated typos variations of an authoritative domain.

%%maybe add some organization here.

\begin{table*}
	\centering
	\def\arraystretch{1}
	{\small
		\caption{Summary of Typo Domain Identification Approaches. {\normalfont $^\dagger$\url{www.MillerSmiles.co.uk} is one of the internet's leading anti-phishing sites, maintaining a massive archive of phishing and identity theft email scams.}}
	\label{tab:tab1}

	\begin{tabular}{|l|l|l|p{8em}|p{8em}|}
		\hline
		\textbf{Approach} & \textbf{Authoritative Domains} & \textbf{Typo Model(s)} & \textbf{Typo Domains Generated} & \textbf{Active Typo Domains}\\
		\hline
		\hline
		\multirow{4}{*}{Wang 2006 \cite{Wang2006}} & Alexa Top 10,000 & (1) Missing-Dot & 10,000 & 51\% (5,094) \\
		& Alexa Top 30 & (1-5) & 3,136 & 71\%(2,233) \\
		& MillerSmiles$^\dagger$ Top 30 & (1-5) & 3,780 & 42\%(1,596) \\
		& Top 50 Children's Sites & (1-5) & 7,094 & 38\%(2,685) \\
		\hline
		Keats 2007 \cite{Keats2007} & Top 2,771 (\textit{Various Sources}) & (1-5) & 1,920,256 & 7\% (127,381)			\\
		\hline
		McAfee Labs 2008 \cite{Edelman2008} & Top 2,000 (\textit{Unknown Source}) & \textit{Unknown} & \textit{Unknown} & 80,000		\\
		\hline
		Banerjee 2008 \cite{Banerjee2008} & Top 900 (\textit{Various Sources}) & (6-8) & $\sim$3 million & 35\% \\	
		\hline
		Moore 2010 \cite{Moore2010} & Alexa Top 3,264 & (1-5) & 1,910,738 & $\sim$49\%(938,000) \\
		\hline
		Szurdi 2014 \cite{Szurdi2014} & Alexa Top 1 million & (1-5) & $\sim$4.7 million & $\sim$20\% \\
		\hline
		Agten 2015 \cite{Agten2015} & Alexa Top 500 & (1-5) & 28,179 & 61\% (17,172) \\		
		\hline			
	\end{tabular}
	}
\end{table*}

\BfPara{Typo-Generation Models}
One of the first and widely cited approaches in this area was introduced by Wang \etal \cite{Wang2006} where given a target domain (e.g. \texttt{www.example.com}), the following five typo-generation models are commonly used:
\begin{enumerate}
	\item \textbf{Missing-dot typos:} this typo happens when the dot following ``www'' is forgotten, e.g., \texttt{wwwexample.com}
	\item \textbf{Character-omission typos:} this typo happens when one character in the original domain name is omitted, e.g., \texttt{www.exmple.com}
	\item \textbf{Character-permutation typos:} this typo happens when two consecutive characters are swapped in the original domain name, e.g., \texttt{www.examlpe.com}
	\item \textbf{Character-substitution typos:} this typo happens when characters are replaced in the original domain name by their adjacent ones on a specific keyboard layout, e.g., \texttt{www.ezample.com}, where ``x'' was replaced by the QWERTY-adjacent character ``z''.
	\item \textbf{Character-duplication typos:} this typo happens when characters are mistakenly typed twice (where they appear once in the original domain name), e.g., {\tt www.exaample.com}
\end{enumerate}

While this previous study presented the first attempt to systematically understand techniques for typosquatting that are most prevalent based on certain usage aspects, later studies looked at exhaustively generating typo domains using other methods. For example, a similar approach in 2008 by Banerjee \etal \cite{Banerjee2008} suggested the following methods for generating typosquatted domains:
\begin{enumerate}\itemsep=-1mm
	\setcounter{enumi}{5}
	\item \textbf{1-mod-inplace:} this typo happens when the typosquatter substitutes a character in the original domain name with all possible alphabet letters.
	\item \textbf{1-mod-deflate:} this typo happens when a typosquatter removes one character from the original domain name (or URL)---and unlike~\cite{Wang2006} where a specific character is considered (e.g., dot), this work systematically considers all possible characters as candidates.
	\item \textbf{1-mod-inflate:} this typo happens when a typosquatter increases the length of a domain name (or URL) by one character. Unlike in~\cite{Wang2006} characters are added based on distance (e.g., using a keyboard layout), this work considers all characters as potential candidates.
\end{enumerate}

Certain aspects of the techniques proposed in~\cite{Banerjee2008} can be viewed as generalization of the techniques proposed in~\cite{Wang2006}. For example, rather than substituting adjacent characters on a keyboard as shown by Wang \etal's fourth model, Banerjee \etal substituted all possible alphabet characters when generating typo domains. In addition, they also experimented with two and three character modifications for their {\bf inplace}, {\bf inflate} and {\bf deflate} schemes thereby generating roughly three million possible typo domain names starting with a corpus of 900 original domain names. 

After probing for the existence of a possible typo domain, Banerjee \etal observed that approximately 99\% of the ``phony'' typosquatted sites they identified utilized a one-character modification of the popular domain names they targeted. Essentially, these are domain names that have a Damerau-Levenshtein distance~\cite{Levenshtein1966} of one from the domains they target. The Damerau-Levenshtein distance is the minimum number of operations needed to transform one string into another, where an operation is defined as an insertion, deletion, or substitution of a single character, or a transposition of two adjacent characters (a generalization of  Hamming distance).

\subsection{Advanced Squatting Techniques}
While the two representative studies discussed in \textsection\ref{sec:models} present examples of systematic typosquatting techniques, other techniques that exploit visual, hardware, and sound similarities have been explored as well. In the following, we review those techniques and their use.

\subsubsection{Homograph Attacks} Per Holgers \etal \cite{Holgers2006}, the homograph attack relies on the visual similarity of letters or strings that might be confused with one another. For example, an attacker can exploit the fact that the lower-case letter `L' ({\sf l}) is visually confusable with the upper-case letter `i' ({\sf I}) in sans-serif fonts and register {\tt www.paypai.com} which targets the popular payment site PayPal.  The end result, in san-serif font, looks very similar to the original: {\sf www.paypal.com} vs {\sf www.paypaI.com}. Alarming as it may seem, the measurement results of Holgers \etal shows that these homograph attacks are rare and not severe in nature. However, these types of attacks may continue to be an attractive choice for would-be cyber-criminals since it can fool most users--as demonstrated in the user study ``Why Phishing Works'' by Dhamija \etal \cite{Dhamija2006}, where 90.9\% of their participants were fooled by such an attack. In that particular case, the researchers generated a phishing website that was an exact replica of the {Bank of the West} homepage that was hosted at {\tt www.bankofthevvest.com}, with two ``v''s instead of a ``w'' in the domain name.

\subsubsection{Bitsquatting} This unique approach to domain squatting was introduced in 2011 by Artem Dinaburg at the BlackHat Security Conference. This technique relies on random bit-errors to redirect connections intended for popular domains \cite{Dinaburg2011}. To test this theory, Dinaburg conducted an experiment and registered 30 {\tt bitsquatted} versions of popular domains (e.g. {\tt www.mic2osoft.com}) and logged all HTTP requests. Much to his surprise, there were a total of 52,317 bitsquat requests from 12,949 unique IP addresses over an eight-month period. Nikiforakis \etal \cite{Nikiforakis2013} studied Dinaburg's findings further and conducted one of the first large-scale analysis of the bitsquatting phenomenon. Their results show that new bitsquatting domains are registered daily and that these attackers monetize their domains through the use of ads, abuse of affiliate programs and even malware installations and distribution. While typosquatting relies on humans to make mistakes, bitsquatting on the other hand relies on computers (hardware) to make mistakes.

\subsubsection{Soundsquatting} Discovered by Nikiforakis \etal \cite{Nikiforakis2014} while researching domain squatting, \textit{soundsquatting} takes advantage of the similarity of words with regard to sound and user confusion on which word represents the desired concept. Unlike typosquatting, soundsquatting does not rely on the typographical mistakes made by users--it is based on \textit{homophones}, which are two words that sound the same but spelled differently (e.g. ``ate'' and ``eight''). To verify how much this soundsquatting technique is used in the wild, Nikiforakis \etal developed a tool to generate possible soundsquatted domains from a list of target domains. Using the Alexa top 10,000 sites, they were able to generate 8,476 soundsquatted domains where 1,823 (21.5\%) of those were already registered. The results presented in \cite{Nikiforakis2014} indeed show that soundsquatting is a viable threat that should be taken into account when defending against domain squatters.

\subsubsection{Typosquatting Cross-site Scripting (TXSS)}
In a study conducted by Nikiforakis \etal \cite{Nikiforakis2012} that examined malicious JavaScript inclusions, they identified a new type of vulnerability that occurs when a web developer mistypes the address of a JavaScript library in their HTML pages or JavaScript code. This simple mistake allows an attacker to register the mistyped domain and easily compromise the site that includes the script. To further explore the impact of this type of attack, the researchers registered a typo variation of a popular JavaScript inclusion domain ({\tt googlesyndicatio.com} vs. {\tt googlesyndication.com}) and observed its traffic: 163,188 unique visitors over the course of 15 days. Nikiforakis \etal argue that the damage of TXSS is much greater than that of typosquatting, since every user visiting the page containing the typo will be exposed to malicious code hosted on the attacker's site.

%\footnote{The service on this domain name is no discontinued and the domain name redirects the {\tt google.com}.}

\subsection{Features of Typosquatted Domains}
In the following, we review features of typosquatted domain names as confirmed by measurements and their evolution over time, including length of domain names (\textsection\ref{sec:length}), popularity of domain names (\textsection\ref{sec:popularity}), popularity of top-level domain (TLD) (\textsection\ref{sec:tld}), and domain landing behavior (\textsection\ref{sec:requests}).

\subsubsection{Domain Name Length}\label{sec:length}
One of the features of domain names investigated for its correlation with typosquatting is their length.
For example, while investigating if domain name length affects the chances of being typosquatted, Banerjee \etal \cite{Banerjee2008} observed that more than 10\% of all possible ``phony'' typosquatted sites registered on the Internet have URLs with less than 10 characters. This fulfills their expectation that typosquatters target domains with shorter names, since popular sites often have short names. 

However, in a contradictory study by Moore and Edelman \cite{Moore2010}, the authors show that no matter the length of the popular domain, typo domains within the {Damerau-Levenshtein} distance of one or adjacent-keyboard distance of one from popular domains were overwhelmingly confirmed as typosquatted. Naturally, we can expect that as the length of domain names increases the probability of it being typosquatted increases since the number of possible typo variations increases. This concept is solidified in the results of the 2015 study by Agten \etal \cite{Agten2015}, which concluded that typosquatters have started targeting longer authoritative domains in the years following 2009, due to the fact that most short typosquatting domains were already in use.

\subsubsection{Domain Name Popularity}\label{sec:popularity}
Another feature of domains names that has been investigated for its correlation with typosquatting is their popularity. 
It is naturally expected that typosquatters will target the most popular domain names to maximize the return on their investment (e.g., the number of visits by unsuspecting users). The results of Banerjee \etal \cite{Banerjee2008} initially suggest that the percentage of active typosquatting domains for a given authoritative domain decreases significantly with the declining popularity. This is in contrast to the results presented by Szurdi \etal \cite{Szurdi2014}, who performed a comprehensive study of typosquatting domain registrations within the {\tt .com} TLD---the largest TLD in the domain name ecosystem. They concluded that 95\% of typo domains target the ``long tail'' of the popularity distribution. The longitudinal study by Agten \etal \cite{Agten2015} also confirms this trend, suggesting a shift in trends and behaviors of typosquatters.

\subsubsection{Effect of the Top-Level Domain}\label{sec:tld}
The popularity of a TLD has been also investigated as a feature for its correlation with typosquatted domain names. 
For example, since the {\tt.com} TLD was introduced as one of the first TLDs when the Domain Name System (DNS) was first implemented in January 1985 \cite{Verisign2015}, it makes up a large portion of the total number of registered domain names (As of June 30, 2015, the total number of registered domain names was 294 million, out of which 117.9 million domain names were registered under {\tt.com}, making up roughly 40\% of the total domain names (\url{http://bit.ly/1VKiMr3})). As such, a majority of the existing studies conducted on typosquatting have only considered domain names in the \texttt{.com} TLD. In their results, Banerjee \etal \cite{Banerjee2008} observed that for nearly a quarter of all initial \texttt{.com} URLs, at least 50\% of all possible phony sites exist; confirming that a domain name ending with {\tt .com} has a high chance of being typosquatted. Interestingly, the results of Agten \etal \cite{Agten2015} finds that certain country-code TLDs ({\tt .uk}, {\tt .jp}, etc.) affect the number of typosquatted domains they contain due to either an unconventional domain dispute policy or domain cost (e.g., cheaper domain names are more likely to be typosquatted).

Additionally, the TLD portion of a domain name may also be a target for exploitation. For example, one {\tt .com} domain may have a malicious {\tt .org} counterpart unbeknownst to the original registrant of the {\tt .com} domain. A noteworthy example of this was mentioned in \cite{Clark2004}, where unsuspecting viewers inadvertently typed {\tt www.whitehouse.com} instead of \texttt{www.whitehouse.gov} and got exposed to questionable contents instead of the official White House website. Banerjee \etal \cite{Banerjee2008} further studied this effect and observed that domains under the {\tt .com} TLD are impersonated primarily in {\tt .biz}, {\tt .net} and {\tt .org} domains, and that domains not registered in the {\tt .com} TLD extension are impersonated primarily in {\tt .com}, {\tt .net} and {\tt .org} domains.

\subsubsection{Probability Models for Domain Landing}\label{sec:requests}
The 2015 study by Khan \etal \cite{Khan2015} introduced a novel approach for detecting typosquatting domains called the \textit{conditional probability model}, which requires a vantage point at the network level to examine DNS and HTTP traffic records. This model identifies domains that have a high proportion of visitors leaving soon after landing on a site (domain name), followed by a visit to a more popular site (domain name) with a similar name. Specifically, they generated pairs of domains ($d_{1}$,$d_{2}$) such that each visit was performed within 33 seconds of each other and the {Damerau-Levenshtein} edit distance between the two domains is one. When dealing with lexically-similar domain pairs, where one of the two domains is unlikely a typo of another, e.g., {\tt nhl.com} and {\tt nfl.com}, the advantage of applying the {conditional probability model} is that it does not correlate such domain pairs. In the results reported by Khan \etal, a request for {\tt nhl.com} is only followed by a load of {\tt nfl.com} .08\% of the time where the reverse rate is even lower at $<0.01\%$. However, they also reported that visits to the site {\tt eba.com} are followed by visits to {\tt ebay.com} 90\% of the time, thus indicating that visits to {\tt eba.com} are likely to be typos.

\section{Monetization Strategies}\label{sec:monetization}
The main drive of typosquatting is monetary in the first place, thus typosquatters employ various techniques to capitalize on their typosquatted domain names and generate revenues. In the following section, we review the various techniques that typosquatters employ to profit from deliberate registrations of typo domain names, including domain name parking (\textsection\ref{sec:parking}), domain name ransoming (\textsection\ref{sec:ransom}), affiliate marketing (\textsection\ref{sec:aff}), hit stealing (\textsection\ref{sec:hit}), and scams (\textsection\ref{sec:scams}).

%\BfPara{Domain Parking} 
\subsection{Domain Parking}\label{sec:parking}
The results of the 2006 study by Wang \etal~\cite{Wang2006} revealed that a large percentage of typo domains they observed were ``parked'', where there was no real content on these pages except for advertisements that were generated by domain parking services. For example, Moore and Edelman's 2010 study \cite{Moore2010} highlighted the case of the typosquatted site {\tt wwwexpendia.com}, which led to a web page that contained a list of sponsored links to travel-related websites. The popular travel site {\tt expedia.com}, the most likely target, was at the top of the list followed by sponsored links to competitors such as {\tt Orbitz.com} and {\tt CheapTickets.com}. In the most recent study on typosquatting conducted by Agten \etal \cite{Agten2015}, domain parking continues to be the most popular scheme chosen by typosquatters.

Domain parking is not limited to benign applications, as show in the previous studies and more recently in~\cite{Metcalf2014}, but may also include malicious behaviors and activities. For example, Alrwais \etal~\cite{Alrwais14} explored the dark side of domain parking, and showed that parked domain names can be actually used for click fraud, traffic stealing, and spam delivery, all of which generate more than 40\% of the revenue for some parking services.

\subsection{Selling and Ransoming Domain Names}\label{sec:ransom}
In addition to being ``parked'' with advertisements, a typosquatted site may have no content other than being advertised as for sale. In the extreme case, these typo domain names may be held for ransom--as in the Zuccarini case highlighted by the 2003 study by Edelman \cite{Edelman}. Edelman found that the vast majority of the typosquatted domain names acquired by the infamous cyber-criminal John Zuccarini were often redirected to websites with sexually-explicit content. For the owners of the authoritative domain names that Zuccarini's typosquatted sites targeted, Edelman argued that having redirections to sexually-explicit content only increased their willingness to pay. Furthermore, Edelman has also pointed out that Zuccarini may have profited from another source of revenue: affiliate marketing programs which we review next.

\subsection{Affiliate Marketing}\label{sec:aff}
These programs are set up by companies to allow third parties to collect commissions on sales or referral fees for redirecting customers to their websites~\cite{chachra2015affiliate}. Such redirection can be for legitimate~\cite{burema2001system} or illicit applications~\cite{levchenko2011click}. For example, the ``Amazon Associates'' program was one of the first online affiliate marketing programs that was launched in 1996 \cite{Amazon}. When ``Associates'' (\ie affiliates) create URL links and potential customers click through those links and buy products from Amazon, the Associates earn referral fees. Typically, these URL links contain unique identifiers to determine which affiliate has forwarded visitors. As Agten \etal point out, many typosquatters abuse such affiliate programs when they redirect visitors to the intended site, collecting referral fees from the authoritative owner for a visit that should have been theirs in the first place \cite{Agten2015}.

\subsection{Hit Stealing}\label{sec:hit}
Not only do typosquatters redirect visitors to their intended websites (for monetary gain), but they can also forward them to websites of competitors. Essentially, these typo domain registrations ``steal'' traffic meant for authoritative domains. The study by Agten \etal \cite{Agten2015} found that this behavior was mostly associated with adult sites (and for spam and click fraud as shown by Alrwais \etal~\cite{Alrwais14}). However, some non-adult sites steal hits from their competitors in situations involving Internet marketing companies who draw traffic to the sites of their customers.

\subsection{Scams}\label{sec:scams}
In this scenario, unsuspecting visitors may fall victim to a scheme that tricks them into divulging personally identifiable information (PII). As reported in \cite{zonealarm}, the typosquatted sites {\tt Wikapedia.com} and {\tt Twtter.com} emulated the real sites ({\tt Wikipedia.org} and {\tt Twitter.com}) and displayed advertisements for contests offering Apple iPads and MacBooks as prizes. Ultimately, users were prompted to enter their credit card number and other sensitive information as part of the contest to claim their prizes.
%%% add a couple of other examples.

\section{Countermeasures}\label{sec:countermeasures}
Countermeasures to typosquatting involve technical and policy-based aspects. Technical aspects to typosquatting, as indicated in the surveyed work in this study, take the identification of typosquatted domains as a first step. Then, the policy-based approach employs legal frameworks to resolve disputes between registrants in case of typosquatted domain names. While the technical aspects are treated at length in~\textsection\ref{sec:anatomy}, in the following we review the policy-based aspects to countermeasures domain name typosquatting.

In November of 1998, the United States Department of Commerce identified a private, non-profit organization called the Internet Corporation for Assigned Names and Numbers (ICANN) as the new entity to oversee the domain name registration system \cite{icann1}. As the Internet rapidly expanded and the number of domain names being registered spiked, cybersquatters and typosquatters alike were quickly snatching up available domain names. In response, the ICANN introduced the Uniform Domain Name Dispute Resolution Policy (UDRP) in late 1999 which states that a domain registrant is required to submit to a mandatory administrative proceeding in the event that a third party (a ``complainant'') disputes such a domain \cite{icann2}. As Moore and Edelman \cite{Moore2010} point out, while the majority of UDRP arbitration proceedings are successful for the complainants, the filing fees can range from \$1,300 to \$4,000---since December 1, 2002 (in use as of March 2016), the world intellectual property organization (WIPO), one of the main arbitration organization assigned by ICANN, charges a tiered fee structure; \$1,500 for up to five domains, and \$2,000 for up to 10 domains in a single complaint reviewed by one panelist, and \$4,000 and \$5,000, respectively, with three panelists (\url{http://bit.ly/1MHwR1c}). While this might not be a lot of money for big companies, it might discourage smaller companies from filing a complaint, especially if targeted by large number of typosquatters/typosquatted domains. 

Since the only remedies available to a complainant in the UDRP are the cancellation or the transfer of the domain name, another alternative became available through legal means: The Anti-cybersquatting Consumer Protection Act (ACPA). As noted in {\em Shields v. Zuccarini}~\cite{2001shields}: ``On November 29, 1999, the ACPA became law, making it illegal for a person to register or to use with the ``bad faith" intent to profit from an Internet domain name that is ``identical or confusingly similar" to the distinctive or famous trademark or Internet domain name of another person or company.'' 

As mentioned earlier, the typosquatters in the Facebook case were found to have violated the ACPA and were ordered to surrender their domain names as well as pay Facebook, netting them a total of \$2.8 million in damages \cite{Roth2013}. While both the UDRP and ACPA can have successful outcomes for the authoritative domain name owners who decide to take the policy intervention route, eliminating the opportunity via defensive registration is perhaps the best strategy. 

Defensive registration is a tactic where companies and trademark owners will deliberately register typo variations of their own domains, keeping it out of the hands of typosquatters and thus redirecting users to the proper domain. Despite this simple strategy, the results of Agten \etal shows that only 156 of the Alexa top 500 have defensive domain registrations, meaning that 344 domains (68.8\%) have no defensive registrations whatsoever \cite{Agten2015}.

In line with defensive registration efforts, various registries offer domain name suggestion and trademark clearinghouse services to reduce the risks associated with typosquatting and typosquatted domain name registration by speculators. For example, ICANN specifies the structure and pricing of trademark clearinghouse, which can be deployed by any interested registry (e.g., Neustar, Nominet, Verisign, etc.)\footnote{\url{http://bit.ly/1Sn77OJ}}. Furthermore, Verisign provides name suggestion services that may include, among their suggestions, typosquatted domains\footnote{\url{http://bit.ly/22yW0Xq}}.

\section{Conclusions and Open Directions}\label{sec:conclusion}\vspace{-2mm}
In this paper we reviewed the landscape of domain name typosquatting and identified techniques used for typosquatting, methods used for their monetization, and countermeasures, including policy-based approaches. 

While the current state-of-the-art highlights the problem, detection techniques, and policy-based approaches, less work is done on the technical front towards defending against this threat. To this end, we foresee a great opportunity in pursuing technical solutions to typosquatting utilizing features and inartistic characteristics of typosquatted domain names. In particular, we are currently pursuing three directions to realize informed solutions to the problem at hand:

\BfPara{End-user feedback} The nature of a domain name is often inferred from certain side channels, as by Khan \etal \cite{Khan2015}, or the type of contents it delivers as per Agten \etal \cite{Agten2015}. However, none of the prior work considered end-users' feedback on the nature of those domain names, and whether they are domains of interest to them. Such ground truth is valuable, and could highlight new trends and features of typosquatted domain names that are not obvious, or captured by the algorithmic models we know so far.  On the other hand, we envision a system that utilizes the well-known features of typosquatted domain names (blacklists of them or new features discovered using users' feedback) to inform users about the risks associated with domain names they are about to visit. Such feedback can be delivered to users in a usable way in the browser.

\BfPara{DNS-level filtering} While the user-centric approach to the problem provides the highest fidelity capturing the users' intent, it does not scale well. To this end, we also foresee a complementary solution that outsources all computations and decisions to the network. For example, based on a fine-grained ground truth of typosquatted domain names, one solution to prevent users from landing on those domain names and exposing themselves to attackers is to implement the blocking of such typosquatted domain names at the DNS level. One realization of such approach to implement the defense using a blacklist as a middlebox in the network. While the approach scales well, and is agnostic to the behavior of users and interaction with the defense system (unlike the end-user feedback based solution), it is also agnostic to the users' intent, and may block domains of interest to users. Furthermore, the system would rely on a blacklist that need to be actively and frequently updated, which comes at cost. 

\BfPara{Up-to-date view via measurements} Many of the studies in the literature concerning the features of typosquatted domain names, their correlation with domain properties, and models for generating them are outdated. Furthermore, some of the recent studies concerning a partial set of those feature refute well-established belief in the earlier studies on this topic. To this end, we foresee a fruitful direction in revising those studies in light of recent datasets (and previously not studied TLDs). In particular, as the use of new generic TLDs (gTLDs) is on the rise, we will extend our measurements to those TLDs in the pursuit for new features for finer understanding of the threat of domain name typosquatting and its evolution.

%\bibliographystyle{abbrv}%  IEEEtran
%\bibliography{IEEEabrv,ref}

\end{document}